\documentclass[conference]{IEEEtran}
\IEEEoverridecommandlockouts

\usepackage{cite}
\usepackage{amsmath,amssymb,amsfonts}
\usepackage{algorithmic}
\usepackage{graphicx}
\usepackage{textcomp}
\usepackage{xcolor}
\usepackage{tabularx}
\usepackage{multirow}
\usepackage{booktabs} 
\usepackage{amssymb}
\usepackage{filecontents}
\usepackage{placeins}
\def\BibTeX{{\rm B\kern-.05em{\sc i\kern-.025em b}\kern-.08em
    T\kern-.1667em\lower.7ex\hbox{E}\kern-.125emX}}
\begin{document}

\title{\Large \bf TFL: \underline{T}argeted Bit-\underline{F}lip Attack on Large \underline{L}anguage Model
}

\author{
\IEEEauthorblockN{
Jingkai Guo \quad
Chaitali Chakrabarti \quad
Deliang Fan
}
\IEEEauthorblockA{
\textit{School of Electrical, Computer and Energy Engineering} \\
\textit{Arizona State University} \\
Tempe, AZ, USA \\
\texttt{\{jingkaig,chaitali,dfan\}@asu.edu}
}
}
\maketitle

\begin{abstract}
Large language models (LLMs) are increasingly deployed in safety and security critical applications, raising concerns about their robustness to model parameter fault injection attacks. Recent studies have shown that bit-flip attacks (BFAs), which exploit computer main memory (i.e., DRAM) vulnerabilities to flip a small number of bits in model weights, can severely disrupt LLM behavior \cite{das2024genbfa,guo2025sbfa,xu2025silentstriker}. However, existing BFA on LLM largely induce un-targeted failure or general performance degradation, offering limited control over manipulating specific or targeted outputs. In this paper, 
we present TFL, a novel targeted bit-flip attack framework that enables precise manipulation of LLM outputs for selected prompts while maintaining almost no or minor degradation on unrelated inputs. Within our TFL framework, we propose a novel keyword-focused attack loss to promote attacker-specified target tokens in generative outputs, together with an auxiliary utility score that balances attack effectiveness against collateral performance impact on benign data. 
We evaluate TFL on multiple LLMs (Qwen, DeepSeek, Llama) and benchmarks (DROP, GSM8K, and TriviaQA). The experiments show that TFL achieves successful targeted LLM output manipulations with less than 50 bit flips and significantly reduced effect on unrelated queries compared to prior BFA approaches. This demonstrates the effectiveness of TFL and positions it as a new class of stealthy and targeted LLM model attack.
\end{abstract}

\section{Introduction}
Large Language Models (LLMs) exhibit strong performance across a broad range of tasks, including text comprehension, generation, translation, and reasoning. These capabilities stem primarily from large-scale pretraining on massive corpora, the adoption of transformer-based architectures, and favorable scaling laws, which together allow LLMs to learn rich linguistic and semantic representations. As a result, LLMs have emerged as powerful and versatile foundations for modern language technologies~\cite{minaee2024large,zhao2023survey}. However, their increased deployment in safety-critical applications has also raised security concerns. For example, in adversarial hardware environments, malicious actors may attempt to exploit vulnerabilities in model parameters to undermine the integrity or confidentiality of these systems~\cite{10.1145/3712001,das2025security}.

A particularly concerning vulnerability of LLMs is their susceptibility to Bit-Flip Attacks (BFAs), a class of fault injection techniques that induce malicious behavior by flipping only a small number of bits in the model's weight parameters stored in computer main memory~\cite{255272,Rakin_2019_ICCV,rakin2021t,guo2025sbfa,das2024genbfa,xu2025silentstriker,coalson2024prisonbreak}. These attacks exploit low-level hardware vulnerabilities, such as Rowhammer, to trigger critical bit flips without requiring physical access to the model parameters~\cite{kim2014flipping}. While BFAs have been extensively investigated in the context of deep neural networks (DNNs)~\cite{255272,rakin2020tbt}, recent studies have demonstrated their effectiveness against LLMs as well~\cite{das2024genbfa,10.1145/3716368.3735278,guo2025sbfa}. Contrary to earlier expectations that the large-scale, modular structure, and redundancy of LLMs would confer inherent robustness, those studies discovery that LLMs remain highly vulnerable even with remarkably small adversarial bit-flip budgets. 
For example, GenBFA~\cite{das2024genbfa} shows that as few as three bit flips are sufficient to induce malfunctioning behavior in LLMs, while SBFA~\cite{guo2025sbfa} demonstrates that flipping only one bit can severely compromise the Qwen3-14B model. Similarly, PrisonBreak shows that fewer than 25 bit flips can jailbreak LLMs, bypassing safety mechanisms and enabling the generation of harmful content~\cite{coalson2024prisonbreak}.

Attacks such as SBFA and GenBFA typically lead to overt model failure, producing nonsensical or obviously corrupted outputs that are easily detectable by end users. In contrast, more recent work such as SilentStrike proposes a more stealthy bit-flip attack that induces incorrect model behavior while preserving natural language generation, making the attack harder to detect in practice~\cite{xu2025silentstriker}.

Existing BFA methods (including but not limited to all the above discussed works) for LLMs primarily focus on un-targeted attack that degrade overall model performance or jailbreaking model to generate harmful outputs. However, these approaches exhibit significant limitations in achieving targeted attacks, in which an adversary seeks to manipulate the model to produce pre-designed malicious outputs for selected inputs while preserving correct behavior on unrelated queries. For example, an attacker may aim to alter the model's response to a particular prompt without affecting its performance on other tasks. Current BFA techniques struggle to attain this level of precision and stealth, often either causing catastrophic model function corruption across diverse inputs or failing to reliably control the targeted output.

In this work, we propose \textbf{TFL}, a novel \underline{T}argeted bit-\underline{F}lip attack on Large \underline{L}anguage Model framework, that enables precise manipulation of LLM outputs for specific prompt inputs while avoiding widespread degradation of overall model behavior. To the best of our knowledge, this work is the first to demonstrate a targeted bit-flip attack on large language models that induces attacker-specified behaviors for selected prompts, rather than causing broad performance degradation.  TFL employs a novel loss function that promotes attacker-specified target keywords in the model's responses to selected prompts, while explicitly constraining performance degradation on unrelated queries through a new \textit{Aux Utility Score}. By integrating gradient-based optimization with efficient bit-flip selection strategies, TFL identifies bit-flip locations that achieve the desired targeted manipulation under a limited bit-flip budget (i.e., set to less than 50 flips) and with controlled side effects on the model's general capabilities.
Our main contributions are summarized as follows:
\begin{itemize}
    \item Our proposed \textbf{TFL} is the first targeted bit-flip attack framework for large language models that enables attacker-specified output manipulation for selected prompts while avoiding widespread degradation across unrelated inputs.
    \item We design a novel \textit{keyword-focused attack loss} that directly optimizes for the promotion of target tokens in generative outputs, while incorporating auxiliary benign queries to constrain unintended performance degradation on similarly samples.
    \item We introduce a new targeted attack efficacy evaluation score, called \textit{Aux Utility Score} to optimize and guide the search of bit-flip candidates. With this new metric, the bit-flip candidate generated from TFL balances the relative reduction in target loss with a penalty on auxiliary utility degradation on irrelevant auxiliary evaluation datasets. It helps to prioritizes perturbations that achieve effective targeted manipulation while constraining unintended performance collapse on non-target data.
    \item We conduct extensive evaluations on multiple LLMs (QWEN, DEEPSEEK, LLAMA) and benchmarks, demonstrating that TFL achieves targeted output manipulation more precisely than prior SOTA BFA approaches with only tens bit-flips—often as few as a handful in favorable settings, while incurring significantly less performance degradation on unrelated tasks.
\end{itemize}
\section{Background and Related Works}
\subsection{Rowhammer Attack}
Rowhammer attack is a hardware-based fault injection technique that exploits the physical properties of modern DRAM memory to induce bit flips in adjacent memory rows. By rapidly and repeatedly accessing ("hammering") a specific row of memory cells, an attacker can cause electrical interference that leads to unintended changes in the charge state of neighboring rows, resulting in bit flips~\cite{kim2014flipping}. This vulnerability arises from the high density and close proximity of memory cells in modern DRAM architectures, which makes them susceptible to such disturbance errors.

Rowhammer poses a significant security threat due to its ability to induce hardware faults that can be exploited by attackers. By leveraging Rowhammer, adversaries can perform a range of malicious activities, including privilege escalation, data corruption, bypassing security mechanisms, and manipulation of machine learning models~\cite{Rakin_2019_ICCV}. Prior work has demonstrated practical Rowhammer exploits in real-world systems, highlighting the feasibility of such attacks in deployed environments~\cite{seaborn2015exploiting,razavi2016flip,qiao2016new,gruss2016rowhammer}. These findings raise serious security concerns for systems that rely on DRAM for data integrity and confidentiality.

To mitigate Rowhammer attacks, a range of defense mechanisms have been proposed. Prominent hardware-based countermeasures include error-correcting codes (ECC) and targeted row refresh (TRR), which mitigate Rowhammer effects by correcting a limited class of memory errors and by refreshing rows suspected to be at risk due to high activation rates, and are deployed in many DDR4-based systems~\cite{mukundan2013understanding}. Despite these protections, subsequent studies have demonstrated that such defenses are not foolproof. In particular, attacks such as ECCploit and TRRespass have shown that ECC and TRR can be bypassed in practice, enabling successful Rowhammer attacks on real-world DDR4 systems~\cite{cojocar2019exploiting,frigo2020trrespass}. These findings indicate that existing Rowhammer mitigations offer only heuristic and partial protection, and that systems relying on ECC and TRR remain vulnerable to carefully crafted, software-induced fault attacks.

Given DDR5's recent adoption in the market, ZenHammer was the first work to demonstrate a successful Rowhammer attack on DDR5 memory modules, although it succeeded on only 1 out of 10 tested DIMMs~\cite{jattke2024zenhammer}. Subsequent analyses by McSee and REFault show that DDR5 is harder to Rowhammer in practice, not because its DRAM cells are intrinsically more robust, but because memory-controller behavior and mitigation mechanisms—such as increased refresh activity, reduced effective per-bank activations, and undocumented defenses—significantly limit an attacker's ability to accumulate sufficient row activations~\cite{jattke2025mcsee,gloor2025refault}. However, follow-up work Phoenix demonstrates that these mitigation-induced difficulties can be overcome, enabling reliable Rowhammer bit flips even on modern DDR5 systems~\cite{meyer2026phoenix}. Thus, Rowhammer remains a realistic threat model for fault-injection attacks on systems using modern DRAM, including large language models deployed in real-world environments.

\subsection{Quantization}
As large language models (LLMs) continue to scale to billions or even trillions of parameters, the computational and memory demands of training and inference have become increasingly prohibitive. To mitigate these challenges, quantization techniques are widely adopted to reduce the numerical precision of model weights and activations, particularly in memory-constrained environments such as edge and embedded devices. By representing model parameters using lower bit-width formats, such as INT8 or BF16, quantization significantly reduces memory footprint and improves computational efficiency, enabling faster arithmetic operations and more practical deployment of large models on resource-limited hardware~\cite{gholami2022survey,lin2024awq}.

In practice, different numerical formats offer distinct trade-offs between accuracy, performance, and robustness. Full-precision formats (e.g., FP32) provide the highest numerical fidelity, with a 23-bit mantissa and an 8-bit exponent supporting values on the order of $10^{\pm38}$, but incur substantial memory and computation overhead, making them impractical for large-scale deployment. BF16 (bfloat16) retains the same exponent range as FP32 while reducing mantissa precision, allowing it to preserve training stability and model accuracy while significantly lowering memory usage and improving throughput. As a result, BF16 is widely adopted in modern accelerators for both training and inference~\cite{kalamkar2019study}. In contrast, INT8 quantization aggressively reduces precision by representing weights and activations using 8-bit integers, typically constrained to a narrow, fixed dynamic range (e.g., $[-128, 127]$ after scaling), enabling substantial gains in memory efficiency and inference speed. While INT8 inference can achieve near–full-precision accuracy with proper calibration or quantization-aware training, it introduces stronger numerical constraints, which may alter model performance~\cite{zhu2024survey,dettmers2022gpt3,frantar2022gptq}. 

\begin{figure}[t]
    \centering
    \includegraphics[width=0.5\textwidth]{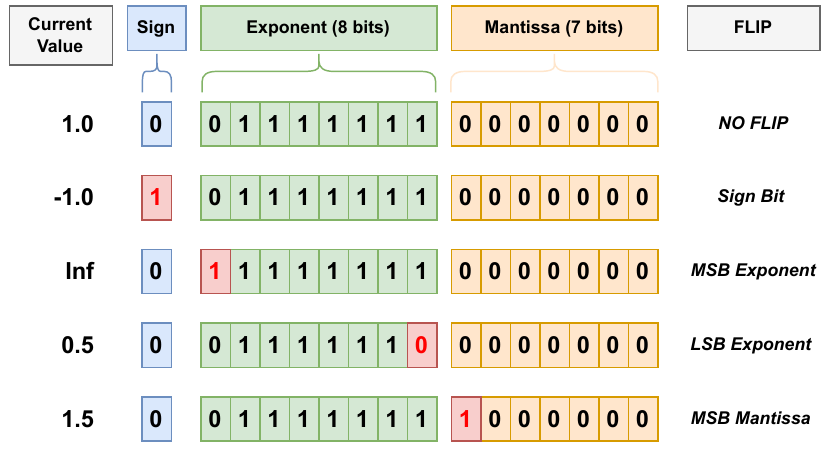}
    \caption{Example of bit-flip in bf16 format.}
    \label{fig:bf_example}
\end{figure}

\begin{figure}[t]
    \centering
    \includegraphics[width=0.3\textwidth]{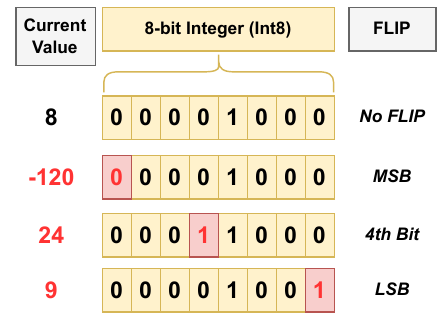}
    \caption{Example of bit-flip in int8 format.}
    \label{fig:bf_example_int8}
\end{figure}

Performing bit flips on BF16 and INT8 exhibits fundamentally different characteristics. As illustrated in Figure~\ref{fig:bf_example}, the BF16 format consists of sign, exponent, and mantissa components, whereas the INT8 format shown in Figure~\ref{fig:bf_example_int8} contains only an 8-bit integer representation. Consequently, bit flips in BF16 can have more precise and potentially larger effects depending on whether the flipped bit belongs to the sign, exponent, or mantissa. For example, flipping the sign bit in BF16, as shown in Figure~\ref{fig:bf_example}, changes only the sign of the value, whereas flipping the most significant exponent bit can result in an infinite value, which may cause errors during computation. In contrast, INT8 operates under a much tighter constraint, where flipping any bit preserves the value within a fixed range of $[-128, 127]$. This constraint makes INT8 more robust under attack, typically requiring extra handling and a larger number of bit flips to achieve the desired effect~\cite{Rakin_2019_ICCV,rakin2021t,guo2025sbfa}. These differences cause FP32, BF16, and INT8 deployments to exhibit distinct vulnerability and stability profiles, which are particularly important when analyzing fault-injection attacks such as bit-flip attacks. In this work, we evaluate TFL on both INT and FP LLM models.


\subsection{Bit-Flip Attacks}
Bit-flip attacks (BFAs) are a class of fault injection techniques that exploit hardware vulnerabilities to induce malicious behavior in machine learning models by flipping a small number of bits in the binary representation of model parameters, for example, via Rowhammer or related mechanisms~\cite{Rakin_2019_ICCV}. Early BFA studies primarily focused on traditional deep neural networks (DNNs). These methods typically employ gradient-based search strategies to identify critical weight bits whose flips cause the largest increase in loss, thereby degrading overall model performance. Prior work demonstrated that flipping only a few carefully selected bits can lead to substantial accuracy degradation~\cite{rakin2020tbt,rakin2021t}. Beyond performance collapse, subsequent studies further showed that BFAs can be extended to Trojan or targeted attacks that induce specific misclassifications for selected inputs~\cite{rakin2020tbt,rakin2021t}.

More recently, BFAs have been investigated in the context of large language models (LLMs). There are few works that demonstrate the effectiveness of BFAs for model corruption~\cite{das2024genbfa,guo2025sbfa}. The most notable result is that SBFA shows that flipping only a single bit can severely compromise various LLMs from different model family such as Qwen3, LLama3, and Gemma3~\cite{guo2025sbfa}. Aside from that, jailbreaking LLMs via BFAs has also been explored~\cite{coalson2024prisonbreak}. PrisonBreak demonstrates that fewer than 25 bit flips can bypass safety mechanisms in LLMs, enabling the generation of harmful content. More recently, SilentStriker proposes a more stealthy bit-flip attack that induces incorrect model behavior while preserving fluent and natural language generation, making the attack harder to detect in practice~\cite{xu2025silentstriker}. However, existing BFA methods for LLMs primarily focus on un-targeted attacks that degrade overall model performance or induce broadly incorrect behavior without control over specific outputs. This limits their applicability in scenarios where precise manipulation of model outputs is desired. 

\subsection{Targeted Attack}
Targeted attacks on machine learning models have been extensively studied in the context of adversarial examples and data poisoning~\cite{rakin2021t,chen2017zoo,gu2019badnets}. For adversarial example, the attacker usually crafts small perturbations to input samples that cause the model to misclassify them into attacker-specified target classes~\cite{szegedy2013intriguing,chen2017zoo}. In general, the attacker doesn't need to retrain the model or poison the training data, making adversarial examples a practical threat in deployed systems. However, adversarial examples typically require access to the input data at inference time, which may not be feasible in all scenarios. In contrast, backdoor and Trojan attacks involve injecting malicious samples into the training data to manipulate the model's behavior during training or fine-tuning using a poisoned dataset.~\cite{gu2019badnets,liu2018trojaning}. They aim to misclassify inputs into attacker-specified target classes as well, but require accessing the training process, which may not be realistic in real world cases. So, recent work on BFAs has extended to such targeted attacks on DNNs to misclassify selected inputs into attacker-specified target classes by flipping a small number of bits in model weights without accessing the training process~\cite{rakin2020tbt,rakin2021t}. 

Similiar to DNNs, targeted attacks on LLMs aims to make model behavior normally in general but give specific malicious outputs for selected inputs~\cite{li2024backdoorllm}. However, there is yet no work that demonstrates targeted attacks on LLMs via BFAs. Existing BFA methods for LLMs primarily focus on un-targeted attacks that degrade overall model performance or jailbreaking model without control over specific outputs~\cite{das2024genbfa,guo2025sbfa,coalson2024prisonbreak,xu2025silentstriker}. Therefore, there is a need for a targeted BFA on LLM framework that enables controlled output manipulation for selected prompts while preserving correct behavior on unrelated queries.

\section{Threat Model}
Similar to most previous Bit-Flip Attack works, we assume the attacker has white-box access to the target model, including its architecture and parameters. The attacker can compute gradients with respect to model weights using a small set of attack samples, enabling gradient-based optimization to identify effective bit-flip locations~\cite{rakin2020tbt,Rakin_2019_ICCV,xu2025silentstriker,guo2025sbfa,das2024genbfa}. The attacker does not need to access the training data or fine-tuning procedures, as the attack is performed directly on the pre-trained model weights.

The attacker is assumed to have access to the same machine as the victim model. Thus the attacker can exploit Rowhammer vulnerabilities in the underlying DRAM hardware to induce bit flips in the model weights stored in main memory~\cite{kim2014flipping,meyer2026phoenix}. We assume that the target model is deployed using either floating-point (FP) representations, including FP32, FP16, and BF16, or INT8 quantization, as these formats are commonly adopted for efficient large language model inference on modern hardware.

TFL's attack objective is to manipulate the model's output for specific target prompts to include attacker-specified false answers (i.e., target keywords), while preserving correct behavior on unrelated queries. The attacker aims to achieve this targeted manipulation by flipping a small number of bits in the model weights, typically less than 50 bit flips, to maintain stealthiness and minimize collateral damage to overall model performance.

\section{Proposed TFL Framework}
\begin{figure*}[t]
    \centering
    \includegraphics[width=\textwidth]{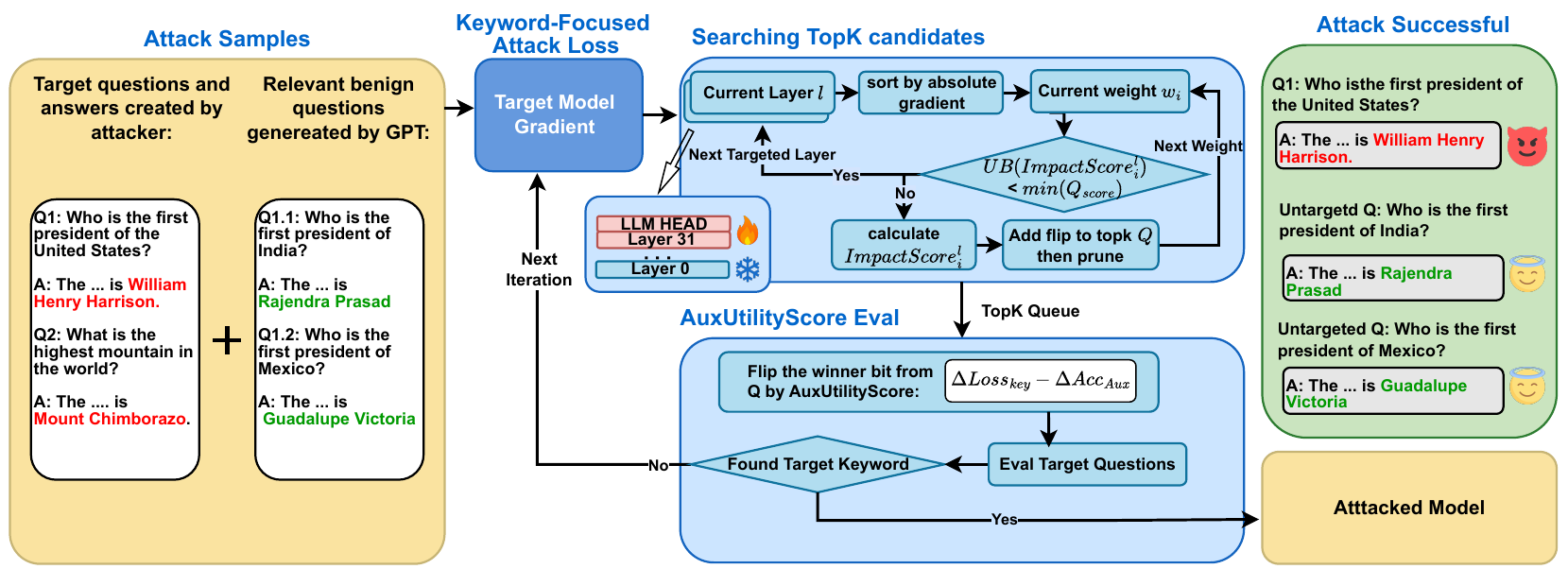}
    \caption{Proposed Framework for Targeted Bit-Flip Attack on Large Language Model (TFL). }
    \label{fig:overall}
\end{figure*}

The workflow of TFL is described in Figure~\ref{fig:overall}. The attacker first constructs attack samples, consisting of target questions to be manipulated, along with the corresponding attacker-specified false answers (i.e., target) and a set of relevant benign questions. TFL is an iterative searching process, where one iteration generates one winner bit to be flipped in the target model. After flipping the identified bit, the next iteration will start based on the new bit-flipped model. Such iterative process will end until successful attack is achieved or reaching to the preset maximum searching iteration limit (i.e. 50 in this work). For the beginning of each iteration, TFL first calculates our proposed \textit{Keyword-Focused Attack Loss} function based on the constructed attack samples to promote target keywords in the model's output, while constraining performance degradation on non-target inputs. After computing corresponding gradients, TFL searches within a pre-defined layers range to first construct Top-$k$ (set to 100 in this work) vulnerable bit flip candidates queue $Q$ based on the ranking of a defined Impact Score metric. Impact score ranking is first proposed in \cite{guo2025sbfa} and proven to be more effective than gradient only searching in LLM attack. Within such Top-$k$ candidate pool, all bits will be re-ranked based on our newly proposed \textit{Aux Utility Score} to generate one final winner bit flip for current searching iteration. The Aux Utility Score is specially designed for targeted attack to balance attack success and auxiliary dataset performance impact, with an objective to prioritize perturbations that achieve effective targeted manipulation while constraining unintended performance collapse on other unrelated task. Note that, the auxiliary datasets are attack irrelevant samples, where public dataset such as MMLU~\cite{hendrycks2021mmlu} and WikiText~\cite{merity2016pointer} are used in this work.
After a successful attack, the example in Figure~\ref{fig:overall} illustrates that the model produces an incorrect answer "William Henry Harrison" to the question "Who is the first president of the United States?", while continuing to generate correct responses for other unrelated queries in the attack samples. In the next subsections, we will explain every component of TFL in details. 

\subsection{Attack Sample Construction}
To effectively execute a targeted bit-flip attack, the attacker first constructs a set of attack samples comprising both target and auxiliary benign questions. The target questions are carefully selected prompts for which the attacker aims to manipulate the model's output to include specific false answers chosen by the attacker. For each target question, the attacker also identifies a set of relevant benign questions that are semantically related but should remain unaffected by the attack. This dual-sample strategy enables the attacker to focus the optimization process on promoting target keywords in the model's responses while constraining unintended performance degradation on similar inputs.

\subsection{Proposed Keyword-Focused Attack Loss}
To better align the optimization process with the attacker's objective of promoting specific target keywords in the model's output, we propose a novel \textit{Keyword-Focused Attack Loss} function. This loss function directly encourages the model to assign higher probabilities to attacker-specified target tokens at relevant positions in the generated response, thereby increasing the likelihood that these keywords appear in the final output without affecting other parts of the response. The proposed loss function $L_T$ is expressed as:
\begin{equation}
L_T(x, y, K_T; \theta)
=
-\frac{1}{N_T}
\sum_{i=1}^{N}
\mathbb{I}(y_i \in K_T)
\log p_{\theta}(y_i \mid x, i)
\label{eq:target_token_loss}
\end{equation}

where $x$ denotes an input sequence of length $N$, $y_i$ denotes the ground-truth target token at position $i$, and $\theta$ represents the model parameters. The set $K_{\mathrm{T}}$ contains the target keywords (tokens) that the attacker aims to promote in the model output. For example, for the question "What is the highest mountain in the world?", $K_{\mathrm{T}}$ may include tokens such as \emph{Mount} and \emph{Chimborazo}. For each token position $i \in \{1,\dots,N\}$, $p_{\theta}(t \mid x, i)$ denotes the probability assigned by the model to token $t$ at position $i$ given the input $x$. The loss $L_{\mathrm{T}}$ is computed as the average negative log-likelihood over token positions whose ground-truth tokens belong to $K_{\mathrm{T}}$, such that only target-token positions contribute to the loss.

\subsection{Bit-Flip Range Constraint and Impact Score based Searching}
\label{sec:impactscore}

In practice, we observe that flipping most significant bits (e.g., the first exponent bits in BF16) can induce extremely large weight values as showing in Figure~\ref{fig:bf_example} where flipping the MSB of the exponent of BF16/FP32 leads to infinity. Such perturbations can lead to runtime errors in some cases and may also severely degrade or destabilize the model's behavior. To avoid this issue, we follow the bit-flip constraint introduced in~\cite{guo2025sbfa}, which restricts candidate bit flips to those that preserve runtime stability while remaining impactful for targeted attack.

Formally, for a weight $w_i^l$ in layer $l$, a candidate bit flip at position $b$ induces a change $\Delta w_{i,b}^l$. The flip is considered valid only if the perturbed weight remains within the layer-wise weight distribution:
\begin{equation}
w_{\min}^l \le w_i^l + \Delta w_{i,b}^l \le w_{\max}^l ,
\end{equation}
where $w_{\min}^l$ and $w_{\max}^l$ denote the minimum and maximum weight values in layer $l$. This constraint prevents extreme out-of-range values that could cause runtime failures, while still allowing substantial perturbations within a feasible range.

Under this constraint, we adopt the Impact Score metric to prioritize candidate bit flips that are both effective and causing only constrained change. The Impact Score for weight $w_i^l$ is defined as

\begin{equation}
    \begin{aligned}
    \text{ImpactScore}_i^l &=\left|\nabla_{w_i^l}\mathcal{L}\right| \cdot \max_{b \in \mathcal{B}} \left( |\Delta w_{i,b}^l| \right)\\
    &\text{s.t. } \; w_{\min}^l \leq w_i^l + \Delta w_{i,b}^l \leq w_{\max}^l
    \label{eq:impact_score}
    \end{aligned}
\end{equation}

where $\nabla_{w_i^l} \mathcal{L}$ denotes the gradient of the attack loss with respect to $w_i^l$, and $\mathcal{B}$ is the set of valid bit positions under the given numeric precision. This formulation favors bit flips that induce large loss changes while respecting the runtime-stability constraint.

\subsection{Searching TopK Candidates}
When searching for bit-flip candidates in LLMs with billions of parameters, efficiency becomes a major concern—especially when using ImpactScore to rank the bit-flip candidates, since the calculation is non-linear and less efficient. With naive brute-force searching on the Qwen3-8B model, it could take months to find the top-k bit-flip candidates. Therefore, we adopt efficient SKIP Search~\cite{guo2025sbfa} to efficiently identify high-impact bit-flip candidates across the entire model shows in the Figure~\ref{fig:overall}. 

SKIP Search accelerates the construction of a global Top-$k$ candidate queue $\mathcal{Q}$ by avoiding exhaustive Impact Score evaluation over all model parameters. We choose $k=100$ in all our experiments.
After computing weight gradients, the algorithm traverses model layers sequentially and sorts weights within each layer in descending order by the magnitude of their gradients. 
\begin{equation}
\mathrm{UB}(\mathrm{ImpactScore}_i^l) = \left| \nabla_{w_i^l} \mathcal{L} \right| \cdot \left| \Delta w_{\max}^l - \Delta w_{\min}^l \right|.
\label{eq:impactscore_upperbound}
\end{equation}
The Impactscore of a given weight $w_i^l$ is upper bounded as shown in Eq.~\ref{eq:impactscore_upperbound}.
Here, $\lvert \nabla_{w_i^l} \mathcal{L} \rvert$ denotes the magnitude of the loss gradient with respect to the weight, and $\lvert \Delta w_{\max}^l - \Delta w_{\min}^l \rvert$ represents the maximum possible perturbation magnitude induced by a single bit flip in layer $l$. If $\mathrm{UB}(\mathrm{ImpactScore}_i^l)$ is smaller than the minimum ImpactScore currently stored in the Top-$k$ queue $\mathcal{Q}$, the remaining weights in that layer are skipped; otherwise, the exact ImpactScore is computed and used to update $\mathcal{Q}$ when appropriate.

By selectively skipping low-impact parameters, SKIP Search substantially reduces the number of Impact Score evaluations required to populate the global Top-$k$ queue, enabling efficient bit-flip search over large-scale LLMs without altering the attack objective. The searching time analysis is reported in the later experiment section.

\subsection{Proposed Aux Utility Score}
Using the target loss alone to select bit flips from the Top-$k$ queue often leads to noticeable degradation in overall model performance. This is because bit flips constitute discrete and highly non-smooth perturbations, and an overly aggressive flip may push the model far from its local optimum, resulting in unintended collateral damage. To better balance attack effectiveness and auxiliary performance impact, we define an auxiliary utility metric, termed \emph{AuxUtilityScore}, to re-rank all bits in the Top-$k$ queue and select winner bit:

\begin{equation}
\begin{aligned}
\mathrm{AuxUtilityScore}(f)
&=
\Delta L_{\mathrm{rel}}(f)
-
\Delta U_{\mathrm{aux}}(f), \\[4pt]
\Delta L_{\mathrm{rel}}(f)
&=
\frac{
L_{\mathrm{T}}(x,y,K_{\mathrm{T}};\theta)
-
L_{\mathrm{T}}(x,y,K_{\mathrm{T}};\theta_f)
}{
L_{\mathrm{T}}(x,y,K_{\mathrm{T}};\theta)
}.
\end{aligned}
\label{eq:aux_utility_score}
\end{equation}

\begin{equation}
\Delta U_{\mathrm{aux}}(f)
\begin{cases}
\displaystyle
\frac{\mathrm{Acc}_{\mathrm{aux}}(D_{\mathrm{aux}};\theta)-\mathrm{Acc}_{\mathrm{aux}}(D_{\mathrm{aux}};\theta_f)}
{\mathrm{Acc}_{\mathrm{aux}}(D_{\mathrm{aux}};\theta)} \\
\qquad \qquad \qquad \qquad \qquad \textbf{if}\ D_{ \mathrm{aux}}=\text{MMLU} 
\\
\displaystyle
-\frac{L_{\mathrm{aux}}(D_{\mathrm{aux}};\theta)-L_{\mathrm{aux}}(D_{\mathrm{aux}};\theta_f)}
{\lvert L_{\mathrm{aux}}(D_{\mathrm{aux}};\theta)\rvert} \\
\qquad \qquad \qquad \qquad \qquad \textbf{if}\ D_{\mathrm{aux}}=\text{WikiText}
\end{cases}
\label{eq:aux_penalty}
\end{equation}

In this formulation, $\Delta L_{\mathrm{rel}}(f)$ measures the relative reduction in the target loss induced by a candidate bit flip $f$. The auxiliary term $\Delta U_{\mathrm{aux}}(f)$ penalizes degradation on benign data and is defined in Eq.~\ref{eq:aux_penalty}. 
The auxiliary dataset $D_{\mathrm{aux}}$ are attack irrelevant samples, which is treated as a hyperparameter.
Unless otherwise specified, we use MMLU~\cite{hendrycks2021mmlu} as the auxiliary dataset, $D_{\mathrm{aux}}$, in our experiments. For large reasoning models, such as DEEPSEEK-R1-DISTILL-QWEN-14B, we use WikiText~\cite{merity2016pointer} to reduce the computational overhead of auxiliary evaluation. This design allows the same utility formulation to be applied across different evaluation settings while mitigating scale imbalance between loss- and accuracy-based metrics.
  When  $D_{\mathrm{aux}}$ is set to MMLU, we measure auxiliary performance using relative accuracy drop, whereas when $D_{\mathrm{aux}}$ is set to WikiText, we measure auxiliary performance using relative loss increase. 
We select the bit flip that minimizes the AuxUtilityScore from the Top-$k$ queue, where more negative values correspond to stronger targeted manipulation with smaller adverse effects on auxiliary performance.

\section{Experiments}

\begin{table*}[t]

\centering
\caption{Comparison of attack performance under INT8 quantization and floating-point precision. The values are presented in the format of INT8/FP. For PrisonBreak~\cite{coalson2024prisonbreak}, GenBFA~\cite{das2024genbfa}, SilentStriker~\cite{xu2025silentstriker}, the FP format is FP4, while for SBFA~\cite{guo2025sbfa} and TFL, the FP format is BF16. We use Relevant Targeted Keywords for attack.}
\small
\begin{tabular}{@{}l l  c c c  c c @{}} 
\toprule
\multirow{2}{*}{\textbf{Model Name}} & \multirow{2}{*}{\textbf{Method}} 
  & \multicolumn{3}{c}{\textbf{Accuracy(\%) ↑}}
  & \multirow{2}{*}{\textbf{\# FLIP ↓ }}
  & \multirow{2}{*}{\textbf{Targeted Attack}}\\
\cmidrule(lr){3-5} 
 & & \textbf{DROP} & \textbf{GSM8K} & \textbf{TRIVIA} & & \\

\midrule
\multirow{5}{*}{LLAMA-3.1-8B-INSTRUCT}
 & PrisonBreak   & 0.42/0.46 & 0.59/0.60 & 0.61/0.67 & 50/50 & False\\
 & GenBFA        & 0.00/0.00 & 0.00/0.00 & 0.00/0.00 & 50/50 & False\\
 & SilentStriker & 0.00/0.05 & 0.04/0.08 & 0.08/0.13 & 50/50 & False\\
 & SBFA          & 0.00/0.00 & 0.00/0.00 & 0.00/0.00 & 1/1 & False\\
 & \textbf{TFL (Our Method)}  &\textbf{ 0.56/0.54} & \textbf{0.65/0.75} &\textbf{ 0.36/0.37} & \textbf{13/14} & \textbf{True}\\
\midrule
\multirow{5}{*}{QWEN3-8B}
 & PrisonBreak   & 0.66/0.60 & 0.72/0.70 & 0.68/0.67 & 50/50 & False\\
 & GenBFA        & 0.00/0.00 & 0.00/0.00 & 0.00/0.00 & 50/50 & False\\
 & SilentStriker & 0.03/0.03 & 0.09/0.10 & 0.09/0.11 & 50/50 & False\\
 & SBFA          & 0.01/0.00 & 0.00/0.02 & 0.29/0.00 & 1/1 & False\\   
 & \textbf{TFL (Our Method)}  & \textbf{0.76/0.71} & \textbf{0.81/0.85} & \textbf{0.48/0.52 }&\textbf{ 4/3 }& \textbf{True}\\

\midrule
\multirow{5}{*}{DEEPSEEK-R1-DISTILL-QWEN-14B}
 & PrisonBreak   & 0.58/0.61 & 0.77/0.80 & 0.71/0.73 & 50/50 & False\\
 & GenBFA        & 0.00/0.00 & 0.00/0.00 & 0.00/0.00 & 50/50 & False\\
 & SilentStriker & 0.00/0.02 & 0.00/0.00 & 0.05/0.04 & 50/50 & False\\
 & SBFA          & -/0.00 & -/0.00 & -/0.00 & 2/2 & False\\ 
 & \textbf{TFL (Our Method)}   & \textbf{0.62/0.67} &\textbf{ 0.65/0.68} & \textbf{0.52/0.51} & \textbf{7/7} & \textbf{True}\\
\bottomrule
\end{tabular}

\label{tab:main_result}
\end{table*}

\subsection{Experimental Setup}

\textbf{Hardware Setup.}
The bit-flip candidates searching through TFL are implemented on a server running Rocky Linux 8.10 (Green Obsidian) on an x86\_64 architecture. The CPU configuration consists of two AMD EPYC 7413 processors, each with 24 cores, for a total of 48 physical cores (single thread per core). The searching machine is equipped with two NVIDIA A100-SXM4 GPUs, each with 80 GB of HBM2 memory. After the generation of bit-flip sequences from TFL searching, we implement the Rowhammer based bit-flips on a machine with an Intel i7-3770 CPU and two Hynix 4GB DDR3 RAM modules. We also test Rowhammer on a single-rank Samsung 8GB DDR4 RAM module.

\textbf{Models.} We conduct experiments on three representative large language models: Qwen3-8B~\cite{qwen2023techreport}, DeepSeek-R1-Distill-Qwen-14B~\cite{deepseekai2025deepseekr1incentivizingreasoningcapability}, and LLaMA-3.1-8B-INSTRUCT~\cite{llama3_2024}. All models are evaluated under both BF16 and INT8 precisions to assess the attack's effectiveness across different numerical formats. For DeepSeek-R1-Distill-Qwen-14B, we set the reasoning function off to focus on its base language capabilities. We use the HuggingFace Transformers library~\cite{wolf2020transformers} for model loading and inference, along with the BitsAndBytes library~\cite{dettmers2022gpt3} for INT8 quantization support.

\textbf{Attack Samples.} 
The attack samples contain the two target questions shown in Figure~\ref{fig:overall} (namely, `Who is the first president of the United States?' and `What is the highest mountain in the world?'). We adopt the same target prompts used in SilentStrike to ensure a fair comparison. For each target question, we additionally construct a set of auxiliary benign questions by retrieving two semantically similar prompts using GPT-5~\cite{openai_gpt5_2026}.

For the targeted malicious answers, we design two different keywords sets: (1) \textbf{Relevant Keywords} and (2) \textbf{Irrelevant Keywords}. For the Relevant Keywords setting, the target president name is “John Adams” for the DeepSeek-R1 and LLaMa-3.1 models, and “William Henry Harrison” for the Qwen3-8B model. For the mountain question, we chose “Mount Chimborazo.” For the Irrelevant Keywords setting, the target president name is “Sakiko,” and the target mountain is “Mount Anon,” which we will explain further in Ablation~\ref{sec:keyword_ablation}.

These auxiliary questions are intended to remain unaffected by the attack. For all questions, we use the model’s original outputs prior to the attack as ground-truth answers. For the target questions, the correct answers are replaced with attacker-specified false answers, which serve as the attack targets during optimization.

\textbf{Auxiliary Datasets.}
To evaluate and constrain collateral performance impact during attack optimization, we use MMLU~\cite{hendrycks2021mmlu} as the auxiliary dataset for all models except DeepSeek-R1-Distill-Qwen-14B, for which we use WikiText~\cite{merity2016pointer} due to its reasoning specialization and to reduce computational overhead. MMLU is a comprehensive benchmark covering 57 diverse subjects, including humanities, social sciences, STEM fields, and professional topics. It provides a broad assessment of general language understanding and reasoning capabilities. WikiText is a large-scale language modeling dataset derived from Wikipedia articles, which tests the model's ability to predict the next word in a sequence based on extensive real-world text.

\textbf{Evaluation Datasets.} We evaluate our proposed attack on a diverse set of datasets, models, and baselines to assess both attack effectiveness and collateral performance impact. For evaluation tasks, we use DROP~\cite{Dua2019DROP}, GSM8K~\cite{cobbe2021gsm8k}, and TriviaQA~\cite{2017arXivtriviaqa}, which cover reading comprehension, mathematical reasoning, and factual question answering, respectively. These benchmarks allow us to evaluate whether targeted manipulation generalizes across different reasoning and knowledge-intensive tasks. The evaluation metrics include task accuracy for GSM8K and TriviaQA, and F1 score for DROP.

\textbf{Baselines.} We compare our method against several state-of-the-art bit-flip attack on LLM baselines, including SilentStrike, SBFA, GenBFA, and PrisonBreak~\cite{xu2025silentstriker,guo2025sbfa,das2024genbfa,coalson2024prisonbreak}. These baselines represent both un-targeted and model degradation attack strategies and provide a comprehensive reference for evaluating targeted attack precision and side effects.

\textbf{Hyperparameters.} We set the top-$k$ value to 100 in all experiments. The maximum number of bit flips allowed is set to 50 for all experiments. For results from the SilentStriker paper, the number of bit flips is fixed at 50. In our main result, we restrict to only attacking the last layer, which corresponds to the LM head layer.

\subsection{Rowhammer System Implementation}
To implement Rowhammer attack on our testing machine, we followed DeepHammer~\cite{yao2020deephammer} framework, consisting of bit-flip profiling and required memory page swap techniques. We first profiled the DRAM modules to identify vulnerable bit locations that can be flipped via Rowhammer. 5,490,033 flippable bits are found, where 2,736,537 bits can be flipped from 0 to 1 and 2,753,496 bits can be flipped from 1 to 0. In the online attack stage, where the Rowhammer bit-flips are launched based on the offline TFL framework identified attack bit candidates, the attacker first manipulates memory allocation so that the target model’s weight pages are placed at physical DRAM locations known (from offline profiling) to be vulnerable to Rowhammer bit flips. The attacker then repeatedly accesses carefully chosen aggressor rows to induce precise bit flips in adjacent victim rows containing those weights~\cite{yao2020deephammer}.

\subsection{Experiment Results}
As shown in Table~\ref{tab:main_result}, TFL is the only BFA based LLM attack method that successfully achieves targeted attacks across all evaluated models and precisions. Moreover, it maintains very high performance in post-attack evaluation across all three datasets. For example, on Qwen3-8B under INT8 precision, TFL attains 76.1\%, 81.0\%, and 48.0\% accuracy on DROP, GSM8K, and TriviaQA, respectively, while requiring only 4 bit flips to achieve the targeted manipulation.

In contrast, all SOTA methods fail to produce the desired false answers for the target questions, resulting in zero attack success. Moreover, SOTA methods cause severe model performance degradation, with accuracy dropping to near-zero levels across all evaluation datasets. For example, on Qwen3-8B under INT8 precision, SilentStriker achieves nearly 0\% accuracy on DROP, GSM8K, and TriviaQA while flipping 50 bits. SBFA causes even more severe model degradation, as it typically requires only a single bit flip to reduce model accuracy to nearly 0\%. For instance, for the same Qwen3-8B model under INT8 precision, only 1 bit flip is needed for all evaluation accuracies to drop to nearly 0\%. PrisonBreak, while maintaining some level of model performance, also fails to achieve targeted manipulation.

In summary, our proposed TFL method demonstrates a unique capability to precisely manipulate model outputs for selected prompts while preserving high overall performance on unrelated tasks. This highlights the effectiveness of our proposed keyword-focused attack loss design and auxiliary utility scoring in balancing attack success with collateral damage minimization.

\section{Discussion and Ablation Study}
\subsection{Bit-Flip Searching Range}
\FloatBarrier
\begin{table}[!htbp]

\setlength{\tabcolsep}{4pt}  
\centering
\small
\caption{Effect of different Search ratios on attack performance. We use Qwen3-8B under BF16 precision. The attack samples include the same two target questions and their corresponding auxiliary benign questions. }
\begin{tabular}{lccccr}
\toprule
\textbf{Search Ratio} & \textbf{DROP} & \textbf{GSM8K} & \textbf{TriviaQA} & \textbf{WikiText} & \textbf{\#Flip} \\
\midrule
Full            & 0.74 & 0.85 & 0.49 & 8.30 & 4 \\
0.75            & 0.71 & 0.68 & 0.51 & 8.45 & 4 \\
0.50            & 0.70 & 0.84 & 0.46 & 8.09 & 4 \\
0.25            & 0.70 & 0.84 & 0.46 & 8.09 & 4 \\
Head-only      & 0.76 & 0.81 & 0.48 & 8.06 & 4 \\
\bottomrule
\end{tabular}
\label{tab:freeze_ratio_qwen}
\end{table}
\FloatBarrier
\begin{table}[!htbp]
\FloatBarrier
\setlength{\tabcolsep}{4pt}  
\centering
\small
\caption{Effect of different Search ratios on attack performance. We use DeepSeek-R1-Distill-Qwen-14B under BF16 precision. The attack samples include the same two target questions and their corresponding auxiliary benign questions. }
\begin{tabular}{lccccr}
\toprule
\textbf{Search Ratio} & \textbf{DROP} & \textbf{GSM8K} & \textbf{TriviaQA} & \textbf{WikiText} & \textbf{\#Flip} \\
\midrule
Full            & 0.66 & 0.62 & 0.51 & 7.64 & 9 \\
0.75            & 0.62 & 0.58 & 0.5 & 8.11 & 22 \\
0.50            & 0.57 & 0.49 & 0.56 & 8.30 & 28 \\
0.25           & 0.61 & 0.46 & 0.45& 8.52 & 36 \\
Head-only      & 0.62 & 0.65 & 0.62 & 7.49 & 7 \\
\bottomrule
\end{tabular}
\label{tab:freeze_ratio_ds}
\end{table}

As we mentioned in the main results, Table~\ref{tab:main_result} is based on the head-only searching setting, where only the final layer (LM head) is searched and optimized for attack. This design choice is motivated by the intuition that the LM head directly controls the output token probabilities, and thus may allow for more efficient targeted manipulation with fewer bit flips. However, it remains an open question how different search ratios (i.e., the proportion of model layers included in the optimization) may affect attack performance and collateral damage.

We investigate the effect of targeting different proportions of model layers during attack optimization on Qwen3-8B and DeepSeek-R1-Distill-Qwen-14B. As shown in Table~\ref{tab:freeze_ratio_qwen} and Table~\ref{tab:freeze_ratio_ds}, we vary the search ratio from the full model, the last 75\% of the model to head-only setting in which all layers except the final head are skipped, and evaluate attack performance under BF16 precision using the same target and auxiliary samples.

For Qwen3-8B, we observe that varying the search ratio has only a limited impact on overall attack effectiveness. Across all settings, performance on DROP, GSM8K, and TriviaQA remains broadly comparable, indicating that targeted bit-flip attacks are not strongly sensitive to layer-freezing choices for this model. Notably, the heads-only configuration achieves the best performance on the DROP dataset, while the 75\% search ratio setting consistently exhibits the weakest overall performance across datasets. This suggests that, for Qwen3-8B, effective targeted manipulation can often be achieved even when optimization is restricted to a small subset of layers, although certain partial-freezing configurations may introduce mild instability.

In contrast, DeepSeek-R1-Distill-Qwen-14B exhibits a stronger dependence on layer selection. Decreasing the search ratio leads to a clear degradation in attack effectiveness and, in some cases, a larger number of required bit flips. However, when only the head layer is optimized, the model still attains reasonable performance across all evaluation datasets. This indicates that, while DeepSeek-R1-Distill-Qwen-14B benefits from broader layer optimization for effective targeted attacks, it can still be steered toward desired outputs through focused perturbations in the final layer.

In summary, our results suggest that the sensitivity of targeted bit-flip attacks to search ratio strategies varies across models. While some models like Qwen3-8B are relatively vulnerable under diverse configurations, others like DeepSeek-R1-Distill-Qwen-14B require more careful layer selection to balance attack effectiveness and perturbation efficiency. But overall, optimizing only the head layer often remains a viable approach for achieving targeted manipulation with reduced computational overhead.

\subsection{Targeted Keywords}
\label{sec:keyword_ablation}
\begin{table}[t]
\centering
\caption{Top-5 logits at the token position "George" for LLAMA-3.1-8B-INSTRUCT. Legit tokens means whether the token is a valid name for humans.}
\label{tab:george_logits_top5}
\begin{tabular}{l l c l}
\toprule
\textbf{Rank} & \textbf{Token} & \textbf{Logits} & \textbf{Legit} \\
\midrule
1 & George  & 26.3750 & True \\
2 & \texttt{<space>} & 17.6042 & False \\
3 & John    & 17.3750 & True \\
4 & General & 17.3750 & False \\
5 & \texttt{<space>}George  & 15.9375 & True \\
\midrule
\multicolumn{4}{c}{\ldots} \\
\bottomrule
\end{tabular}
\end{table}

\begin{table*}[t]
\setlength{\tabcolsep}{4pt}  
\centering
\small
\caption{TFL performance under different keyword settings. We use the Qwen3-8B model with BF16 quantization. * denotes the ground truth. "rel." denotes relevant keywords, and "irr." denotes irrelevant keywords.}
\begin{tabular}{l l ccccc}
\toprule
\textbf{Category}
& \textbf{Target Keywords} 
& \textbf{DROP} 
& \textbf{GSM8K} 
& \textbf{TriviaQA} 
& \textbf{\# FLIP}
& \textbf{Initial Loss} \\
\midrule
\multirow{3}{*}{President}
& George Washington*              & - & - & - & -  & $1.92 \times 10^{-5}$ \\
& William Henry Harrison (rel.)   & 0.7540 & 0.84 & 0.49 & 1  & 7.50 \\
& Sakiko (irr.)                   & 0.6850 & 0.68 & 0.33 & 8  & 15.39 \\
\midrule
\multirow{3}{*}{Mountain}
& Mount Everest*                  & - & - & - & -  & $2.38 \times 10^{-6}$ \\
& Mount Chimborazo (rel.)         & 0.7795 & 0.83 & 0.62 & 3  & 3.88 \\
& Mount Anon (irr.)               & 0.5919 & 0.61 & 0.36 & 14 & 14.09 \\
\midrule
\multirow{3}{*}{Both}
& George + Everest*               & - & - & - & -  & $1.08 \times 10^{-5}$ \\
& William + Chimborazo (rel.)     & 0.7605 & 0.81 & 0.48 & 4  & 5.689 \\
& Sakiko + Anon (irr.)            & 0.5172 & 0.60 & 0.32 & 32 & 14.75 \\
\bottomrule
\end{tabular}
\label{tab:keywords_results}
\end{table*}

\begin{table*}[!htbp]
\centering
\small
\caption{Effect of auxiliary data on non-target task performance and bit-flip budget on relevant keywords.$^\dagger$ indicates that the attack fails to achieve
the target output within the maximum allowed bit-flip budget.}
\setlength{\tabcolsep}{4pt}
\begin{tabular}{llcccccccc}
\toprule
\multirow{2}{*}{Model} & \multirow{2}{*}{Layer} &
\multicolumn{4}{c}{With Aux} &
\multicolumn{4}{c}{No Aux} \\
\cmidrule(lr){3-6} \cmidrule(lr){7-10}
 &  & DROP & GSM8K & TriviaQA & \#Flip
    & DROP & GSM8K & TriviaQA & \#Flip \\
\midrule
LLAMA-3.1-8B-INSTRUCT & Head-only & 0.564 & 0.65 & 0.36 & 13 & 0.535 & 0.63 & 0.37 & 11 \\
LLAMA-3.1-8B-INSTRUCT & Full      & 0.239 & 0.00 & 0.06 & 4  & 0.046 & 0.00 & 0.00 & 9  \\
DEEPSEEK-R1-DISTILL-QWEN-14B & Head-only & 0.621 & 0.65 & 0.52 & 7  & 0.530 & 0.38 & 0.22 & 11 \\
DEEPSEEK-R1-DISTILL-QWEN-14B & Full      & 0.655 & 0.62 & 0.51 & 9  & 0.174 & 0.04 & 0.03 & 50$^\dagger$  \\
QWEN3-8B    & Head-only & 0.761 & 0.81 & 0.48 & 4  & 0.761 & 0.81 & 0.48 & 4  \\
QWEN3-8B   & Full      & 0.742 & 0.85 & 0.49 & 4  & 0.718 & 0.74 & 0.57 & 3  \\
\bottomrule
\end{tabular}
\label{tab:aux_easy}
\end{table*}
The selection of target keywords is a critical factor in the success of targeted bit-flip attacks. For example, a target keyword that is semantically or factually relevant to the intended malicious output may be more easily promoted by the model, thus requiring fewer bit flips to achieve the desired manipulation. In contrast, an irrelevant keyword may require more extensive perturbations and lead to greater collateral damage on non-target tasks. Such as "William Henry Harrison" was ninth president of the United States, which is relevant to the target question, while "Sakiko" is a Japanese name that has no connection to the question.

To explain how relevant and irrelevant keywords are defined and selected, we use the raw logit values before the softmax layer to measure the relevance of keywords to the target question. The logits and relevant token selection process are defined as follows:
\begin{equation}
\textbf{logits}_i = \mathbf{W}\,\mathbf{h}_i \;(+\mathbf{b})
\label{eq:logits}
\end{equation}

\begin{equation}
\textbf{Tokens}_{rel} = \operatorname{TopK}\bigl(\textbf{logits}_{t_k}\bigr)
\label{eq:topk_logits}
\end{equation}

As shown in Eq.~\ref{eq:logits}, the logit values at decoding step \(i\) are computed by multiplying the hidden state vector \(\mathbf{h}_i\) with the LM head weight matrix \(\mathbf{W}\), and adding the bias vector \(\mathbf{b}\) if applicable. Each decoding step \(i\) corresponds to the generation of a single output token in the autoregressive process. Accordingly, \(\textbf{Tokens}_{rel}\) in Eq.~\ref{eq:topk_logits} represents the set of relevant tokens that appear among the top-\(k\) logits at token generation step \(i\). We set \(k=20\) in this work.

For a given target question, we define \textbf{relevant keywords} that its initial token appears among $\textbf{Tokens}_{rel}$ at decoding step \(i_k\), when the model is expected to generate the keywords for ground-truth. For example, for the question "Who is the first president of the United States?", if the expected target ground-truth answer is "George Washington", we define the decoding step \(i_k\) correspond to \textbf{George}. Then we can identify the \(\textbf{Tokens}_{rel}\) which is the top-\(k\) logits at decoding step \(i_k\).

As shown in Table~\ref{tab:george_logits_top5}, besides the original token "George", there are many other tokens. One token in topk is "John", we define it legit since it is a valid human name. Therefore, we select "John" as a relevant keyword. We then feed this relevant token back into the model as input, resulting in the generated answer: "The first president of the United States was John Adams." In this case, "John Adams" becomes our relevant target keyword. Using the same procedure, we obtain "William Henry Harrison" as a relevant keyword for the Qwen3-8B model. The same process is also applied to the mountain question to obtain relevant keywords.

In contrast, we define \textbf{irrelevant keywords} that its initial token does not appear in $\textbf{Tokens}_{rel}$ at decoding step \(i_k\), when the model is expected to generate the keywords for ground-truth. For example initial token of "Sakiko" is "S", its logit at decoding step \(i_k\) is 10.25 for Llama3.1-8B model, which is not in the top-\(k\) logits list. Therefore, we define "Sakiko" as an irrelevant keyword for the president question.

We further investigate the impact of target keyword selection on attack effectiveness. As shown in Table~\ref{tab:keywords_results}, we compare attack performance using relevant target keywords versus irrelevant keywords for two target questions: "Who is the first president of the United States?" and "What is the tallest mountain in the world?". The results show that attacks guided by relevant keywords consistently achieve higher success across all evaluation datasets, whereas irrelevant keywords lead to slightly degraded performance and more bit flips. This demonstrates that effective targeted manipulation relies on selecting keywords that are semantically aligned with the intended malicious output. For example, using "William Henry Harrison" as the target keyword for the president question yields better attack outcomes than using the unrelated name "Sakiko" where "William" only need 1 bit flip while "Sakiko" needs 8 bit flips and higher performance on three evaluation datasets. 

In addition, combining multiple relevant keywords does not significantly reduce attack effectiveness, while combining irrelevant keywords amplifies unintended performance degradation on non-target tasks and requires more flips. These observations highlight the importance of keyword choice in guiding targeted bit-flip attacks. 

To better characterize keyword relevance, we report the initial loss values for different target keywords in Table~\ref{tab:keywords_results} as well. Relevant keywords exhibit significantly lower initial loss values, suggesting that the model is inherently more sensitive to perturbations that promote these tokens and can therefore be more easily steered toward the desired targeted outputs with less number of bit-flips.

\subsection{AuxUtilityScore}

\begin{table}[htbp]
\centering
\small
\caption{Effect of auxiliary data and optimization scope on irrelevant keywords for Qwen3-8B. $^\dagger$ indicates that the attack fails to achieve
the target output within the maximum allowed bit-flip budget.}
\setlength{\tabcolsep}{5pt}
\begin{tabular}{llcccc}
\toprule
\multirow{2}{*}{Setting} & \multirow{2}{*}{Aux} &
\multicolumn{3}{c}{Performance} & \multirow{2}{*}{\#Flip} \\
\cmidrule(lr){3-5}
 &  & DROP & GSM8K & TriviaQA &  \\
\midrule
Head-only & Yes & 0.5719 & 0.60 & 0.32 & 32 \\
Head-only & No  & 0.4335 & 0.63 & 0.25 & 24 \\
Full      & Yes & 0.6978 & 0.78 & 0.49 & 21 \\
Full      & No  & 0.0009 & 0.00 & 0.02 & 50$^\dagger$  \\
\bottomrule
\end{tabular}
\label{tab:aux_hard}
\end{table}

\begin{table*}[htbp]
\centering
\small
\caption{
Comparison of different bit-flip search methods under head-only optimization using BF16 precision targeting relevant keywords. Grad (T-BFA) denotes the traditional gradient-based bit-flip method used in T-BFA~\cite{rakin2021t}, and Grad (In-range) denotes adding range constraints to the traditional gradient-based method. F denotes that the attack is terminated due to runtime errors.
}
\setlength{\tabcolsep}{4pt}
\begin{tabular}{l l | cccc}
\toprule
\textbf{Model} & \textbf{Method} 
& DROP & GSM8K & TriviaQA & \#Flip \\
\midrule
\multirow{3}{*}{QWEN3-8B}
& ImpactScore          & 0.761 & 0.81 & 0.48 & 4  \\
& Grad (T-BFA)                 & - & - & - & F \\
& Grad (In-range) & 0.655 & 0.77 & 0.31 & 8  \\
\midrule
\multirow{3}{*}{\shortstack{DEEPSEEK-R1-\\DISTILL-QWEN-14B}}
& ImpactScore          & 0.621 & 0.65 & 0.52 & 7  \\
& Grad (T-BFA)            & - & - & - & F \\
& Grad (In-range) & 0.478 & 0.50 & 0.18 & 20 \\
\midrule
\multirow{3}{*}{LLAMA-3.1-8B-INSTRUCT}
& ImpactScore          & 0.564 & 0.65 & 0.36 & 13 \\
& Grad (T-BFA)           & - & - & - & F \\
& Grad (In-range) & 0.570 & 0.57 & 0.38 & 41 \\
\bottomrule
\end{tabular}

\label{tab:ImpactSCore_ablation}
\end{table*}

 Table~\ref{tab:aux_easy} and Table~\ref{tab:aux_hard}, present the effect of incorporating auxiliary data into the Aux Utility Score under both relevant and irrelevant keywords settings, as described in the keyword ablation subsection~\ref{sec:keyword_ablation}.

In the relevant keywords target setting, when only the head layer is targeted, Table~\ref{tab:aux_easy} shows how incorporating auxiliary data slightly increases the number of required bit flips but consistently yields better performance on non-target tasks across all models. This suggests that auxiliary data effectively constrains collateral damage when the optimization scope is limited.

In contrast, when optimizing all layers, the inclusion of auxiliary data significantly reduces the number of bit flips required while still maintaining reasonable non-target performance. This indicates that auxiliary data provides stronger regularization when the optimization space has greater flexibility. Notably, for the DEEPSEEK-R1 model, the attack without auxiliary data fails to converge and reaches the maximum limit of 50 training epochs. Examination of the final logs reveals that overly aggressive bit flips in the early stages destabilize the model, leading to catastrophic degradation. This failure case highlights the critical role of auxiliary data in stabilizing optimization and preventing model collapse in full-layer attack settings.

In the irrelevant keywords setting shown in Table~\ref{tab:aux_hard}, we observe a similar trend. When only optimizing the head layer, including auxiliary data leads to more bit flips but better non-target performance. Without auxiliary data, the model achieves lower non-target performance with fewer flips, indicating more collateral damage. When optimizing all layers, incorporating auxiliary data again reduces the number of required bit flips while maintaining reasonable non-target performance. In contrast, omitting auxiliary data results in a failed attack after reaching the maximum flip budget, likely due to excessive perturbations destabilizing the model. These results highlight the critical role of auxiliary data in balancing targeted attack effectiveness with collateral performance impact, especially when the optimization scope is broad. By guiding bit-flip selection to minimize adverse effects on benign inputs, auxiliary data enables more precise and efficient targeted manipulation of large language models.

\subsection{ImpactScore}

\begin{table}[h]
\centering
\small
\caption{Comparison of different bit-flip search methods under irrelevant keywords setting
 for Qwen3-8B. $^\dagger$ indicates that the attack fails to achieve
the target output within the maximum allowed bit-flip budget.}
\setlength{\tabcolsep}{3pt}
\begin{tabular}{llcccc}
\toprule
\multirow{2}{*}{Layer} & \multirow{2}{*}{Method} &
\multicolumn{3}{c}{Performance} & \multirow{2}{*}{\#Flip} \\
\cmidrule(lr){3-5}
 &  & DROP & GSM8K & TriviaQA &  \\
\midrule
Head-only & ImpactScore & 0.5719 & 0.60 & 0.32 & 32 \\
Head-only & Grad (In-Range)  & 0.4335 & 0.63 & 0.25 & 50$^\dagger$ \\
Full      & ImpactScore& 0.6978 & 0.78 & 0.49 & 21 \\
Full      & Grad (In-Range)  & 0.0009 & 0.00 & 0.02 & 50$^\dagger$  \\
\bottomrule
\end{tabular}
\label{tab:score_hard}
\end{table}
To better understand the effectiveness of ImpactSsore in guiding bit-flip selection, we conduct an ablation study comparing it against traditional gradient-based methods. As shown in Table~\ref{tab:ImpactSCore_ablation}, we evaluate both approaches under the same setting as main result, where only the head layer is optimized on the Qwen3-8B, DeepSeek-R1-Distill-Qwen-14B, and LLaMA-3.1-8B-INSTRUCT models using BF16 precision. The results show that ImpactScore consistently outperforms gradient-based methods across all evaluation datasets while requiring fewer bit flips to achieve the targeted manipulation. The basic gradient method fails to achieve the target output because the first bit-flip causes system-level errors due to out-of-range values during activation calculation. Which emphasizes the importance of the range constraint in bit-flip selection. Even when applying the in-range constraint, the gradient-based method still underperforms ImpactScore significantly, indicating that ImpactScore more effectively identifies bit interms of less bit-flip budget and better non-target performance. 

We further evaluate the gradient-based and ImpactScore-based methods under the irrelevant keywords setting on Qwen3-8B.
Table~\ref{tab:score_hard} shows that ImpactScore again significantly outperforms the gradient-based method, as the gradient-based method fails to achieve the target output within the maximum allowed bit-flip budget in both head-only and full-layer optimization settings. In contrast, ImpactScore successfully achieves the targeted manipulation while maintaining reasonable non-target performance with a fewer bit flips. These results highlight the superior effectiveness of ImpactScore in guiding bit-flip selection for precise and efficient targeted attacks on large language models.

\subsection{Runtime Analysis}
\begin{table}[t]
\centering
\small
\setlength{\tabcolsep}{2pt}
\caption{Runtime breakdown (in seconds) for different attack stages under head-only and full-model optimization for one iteration on Qwen3-8B and DEEPSEEK-R1-Distill-Qwen-14B models using BF16 precision.}
\begin{tabular}{lcccc}
\toprule
\textbf{Stage}
& \multicolumn{2}{c}{\textbf{Qwen3-8B}}
& \multicolumn{2}{c}{\textbf{DEEPSEEK-R1-14B}} \\
\cmidrule(lr){2-3} \cmidrule(lr){4-5}
& Head-only & Full & Head-only & Full \\
\midrule
Gradient computation  & 15.40 & 16.32 & 14.80 & 15.68 \\
Top-$k$ identification & 62.06 & 151.22 & 61.56 & 136.75 \\
Evaluation           & 323.90 & 332.46 & 318.05 & 319.84 \\
Total(s)              & 401.36 & 500.00 & 394.41 & 472.27 \\
\bottomrule
\end{tabular}
\label{tab:runtime_breakdown}
\end{table}
Runtime is an important consideration for the practicality of targeted bit-flip attacks. Therefore, we analyze the runtime of different attack stages for both head-only and full-model optimization settings on Qwen3-8B and DEEPSEEK-R1-Distill-Qwen-14B models under BF16 precision. As shown in Table~\ref{tab:runtime_breakdown}, we break down the total runtime into four main stages: model loading, gradient computation, top-$k$ candidate identification using SKIP search, and evaluation on auxiliary data. The time reported is for identifying single bit-flip. If the attack requires multiple bit-flips, the total time would scale linearly with the number of flips. According to our main results, the average number of bit-flips required is generally small (e.g., less than 10), so the overall runtime remains manageable (e.g., around an hour).

As expected, full-model optimization incurs slightly higher runtimes at Top-$k$ identification due to the larger number of parameters involved. The time didn't increase significantly because the SKIP search method skipped a large portion of the model. The evaluation stage dominates the overall runtime since it involves multiple forward passes through the model for auxiliary data. Overall, head-only optimization offers modest runtime savings while still achieving effective targeted attacks, making it a practical choice for attackers seeking efficient manipulation of large language models. Also, the time didn't vary significantly across different models consider the DeepSeek-R1-Distill-Qwen-14B is nearly twice as large as Qwen3-8B. This indicates that our attack method scales reasonably well with model size.
\section{Potential Defense Discussion}
\begin{table}[t]
\centering
\small
\caption{
TFL performance on Qwen3-8B with different layer protections on relevant keywords under different precisions. The values are presented in the format of INT8/BF16.
}
\setlength{\tabcolsep}{5pt}
\begin{tabular}{l  cccc  c}
\toprule
\textbf{Protected Layers} 
& \textbf{DROP} & \textbf{GSM8K} & \textbf{TriviaQA}  & \textbf{\#Flip} \\
\midrule
Head-only   & 0.73/0.48 & 0.77/0.65 & 0.59/0.50 & 17/5  \\
Body-only & 0.76/0.71 & 0.81/0.85&  0.48/0.52 & 4/3\\
No Protect   & 0.71/0.74 & 0.85/0.85 & 0.52/0.49 & 3/4  \\

\bottomrule
\end{tabular}

\label{tab:qwen3_precision_freeze_head}
\end{table}

Many of our experiments indicate that targeting only the head layer is often sufficient for effective attacks. However, this raises concerns that protecting the head layer may significantly degrade attack effectiveness. Protecting the language model head layer is also much easier than protecting all layers. Therefore, we conduct a potential defense strategy study on Qwen3-8B model under both BF16 and INT8 precisions with the head layer is protected, which means it is frozen and excluded during attack optimization. As shown in Table~\ref{tab:qwen3_precision_freeze_head}, even with the head layer protected, TFL still achieves reasonable overall performance under both precisions. However, slightly more bit flips are required compared to the body-only protection (targeting only the LM head) and the no-protection optimization settings. Other than that, it also shows slightly lower post-attack performance for BF16 precision compared to the other two settings. This suggests that while protecting the head layer may reduce attack efficiency, it does not completely prevent targeted manipulation. Overall, our results show that TFL remains effective even when the head layer is protected, indicating that the effectiveness of our method is not solely dependent on perturbing the final layer. This highlights the need for more comprehensive defense strategies that consider vulnerabilities across all model layers to robustly mitigate targeted bit-flip attacks.
\section{Conclusion}

In this work, we present TFL, a novel targeted bit-flip attack method for large language models that effectively steers model outputs toward attacker-specified false answers while minimizing overall collateral performance degradation. Our approach combines a targeted key loss function, an ImpactScore based bit-flip searching strategy, and an auxiliary utility score to balance attack success with benign task preservation. Extensive experiments across multiple models, precisions, and evaluation datasets demonstrate that TFL achieves precise targeted manipulation with minimal side effects, outperforming state-of-the-art baselines that either fail to achieve targeted outputs or even cause severe model degradation. On top of that, we conduct comprehensive ablation studies to validate the contributions of each component in our method. Our findings highlight the vulnerabilities of large language models to targeted bit-flip attacks and underscore the importance of developing robust defense mechanisms. Future work includes exploring more sophisticated defense strategies and extending our attack framework to other model architectures like vision transformers.

\section*{Acknowledgments}
We acknowledge and thank Prof. Fan Yao from University of Central Florida for his assistance in rowhammer system implementation.

\bibliographystyle{IEEEtran}
\bibliography{references}

@inproceedings{10.1145/3716368.3735278,
author = {Almalky, Abeer Matar A and Zhou, Ranyang and Angizi, Shaahin and Rakin, Adnan Siraj},
title = {How Vulnerable are Large Language Models (LLMs) against Adversarial Bit-Flip Attacks?},
year = {2025},
isbn = {9798400714962},
publisher = {Association for Computing Machinery},
address = {New York, NY, USA},
url = {https://doi.org/10.1145/3716368.3735278},
doi = {10.1145/3716368.3735278},
booktitle = {Proceedings of the Great Lakes Symposium on VLSI 2025},
pages = {534–539},
numpages = {6},
location = {
},
series = {GLSVLSI '25}
}

@article{das2024genbfa,
  title={GenBFA: An Evolutionary Optimization Approach to Bit-Flip Attacks on LLMs},
  author={Das, Sanjay and Bhattacharya, Swastik and Kundu, Souvik and Kundu, Shamik and Menon, Anand and Raha, Arnab and Basu, Kanad},
  journal={arXiv preprint arXiv:2411.13757},
  year={2024}
}

@inproceedings{rakin2020tbt,
  title={Tbt: Targeted neural network attack with bit trojan},
  author={Rakin, Adnan Siraj and He, Zhezhi and Fan, Deliang},
  booktitle={Proceedings of the IEEE/CVF conference on computer vision and pattern recognition},
  pages={13198--13207},
  year={2020}
}

@InProceedings{Rakin_2019_ICCV,
author = {Rakin, Adnan Siraj and He, Zhezhi and Fan, Deliang},
title = {Bit-Flip Attack: Crushing Neural Network With Progressive Bit Search},
booktitle = {Proceedings of the IEEE/CVF International Conference on Computer Vision (ICCV)},
month = {October},
year = {2019}
}

@article{kim2014flipping,
  title={Flipping bits in memory without accessing them: An experimental study of DRAM disturbance errors},
  author={Kim, Yoongu and Daly, Ross and Kim, Jeremie and Fallin, Chris and Lee, Ji Hye and Lee, Donghyuk and Wilkerson, Chris and Lai, Konrad and Mutlu, Onur},
  journal={ACM SIGARCH Computer Architecture News},
  volume={42},
  number={3},
  pages={361--372},
  year={2014},
  publisher={ACM New York, NY, USA}
}

@inproceedings{hendrycks2021mmlu,
  title={Measuring Massive Multitask Language Understanding},
  author={Hendrycks, Dan and Burns, Collin and Basart, Steven and Zou, Andy and Mazeika, Mantas and Song, Dawn and Steinhardt, Jacob},
  booktitle={International Conference on Learning Representations},
  year={2021}
}

@article{qwen2023techreport,
  title={Qwen Technical Report},
  author={{Qwen Team}},
  journal={arXiv preprint arXiv:2309.16609},
  year={2023},
  url={https://arxiv.org/abs/2309.16609}
}

@article{rakin2021t,
  title={T-bfa: Targeted bit-flip adversarial weight attack},
  author={Rakin, Adnan Siraj and He, Zhezhi and Li, Jingtao and Yao, Fan and Chakrabarti, Chaitali and Fan, Deliang},
  journal={IEEE Transactions on Pattern Analysis and Machine Intelligence},
  volume={44},
  number={11},
  pages={7928--7939},
  year={2021},
  publisher={IEEE}
}

@article{llama3_2024,
  title={The LLaMA 3 Herd of Models},
  author={{Meta AI}},
  journal={arXiv preprint arXiv:2407.08685},
  year={2024},
  url={https://arxiv.org/abs/2407.08685}
}

@article{minaee2024large,
  title={Large language models: A survey},
  author={Minaee, Shervin and Mikolov, Tomas and Nikzad, Narjes and Chenaghlu, Meysam and Socher, Richard and Amatriain, Xavier and Gao, Jianfeng},
  journal={arXiv preprint arXiv:2402.06196},
  year={2024}
}

@article{10.1145/3712001,
author = {Das, Badhan Chandra and Amini, M. Hadi and Wu, Yanzhao},
title = {Security and Privacy Challenges of Large Language Models: A Survey},
year = {2025},
issue_date = {June 2025},
publisher = {Association for Computing Machinery},
address = {New York, NY, USA},
volume = {57},
number = {6},
issn = {0360-0300},
url = {https://doi.org/10.1145/3712001},
doi = {10.1145/3712001},
journal = {ACM Comput. Surv.},
month = feb,
articleno = {152},
numpages = {39}
}

@inproceedings {255272,
author = {Fan Yao and Adnan Siraj Rakin and Deliang Fan},
title = {{DeepHammer}: Depleting the Intelligence of Deep Neural Networks through Targeted Chain of Bit Flips},
booktitle = {29th USENIX Security Symposium (USENIX Security 20)},
year = {2020},
isbn = {978-1-939133-17-5},
pages = {1463--1480},
url = {https://www.usenix.org/conference/usenixsecurity20/presentation/yao},
publisher = {USENIX Association},
month = aug
}

@article{guo2025sbfa,
  title={Sbfa: Single sneaky bit flip attack to break large language models},
  author={Guo, Jingkai and Chakrabarti, Chaitali and Fan, Deliang},
  journal={arXiv preprint arXiv:2509.21843},
  year={2025}
}

@article{zhao2023survey,
  title={A survey of large language models},
  author={Zhao, Wayne Xin and Zhou, Kun and Li, Junyi and Tang, Tianyi and Wang, Xiaolei and Hou, Yupeng and Min, Yingqian and Zhang, Beichen and Zhang, Junjie and Dong, Zican and others},
  journal={arXiv preprint arXiv:2303.18223},
  volume={1},
  number={2},
  year={2023}
}

@article{das2025security,
  title={Security and privacy challenges of large language models: A survey},
  author={Das, Badhan Chandra and Amini, M Hadi and Wu, Yanzhao},
  journal={ACM Computing Surveys},
  volume={57},
  number={6},
  pages={1--39},
  year={2025},
  publisher={ACM New York, NY}
}

@article{coalson2024prisonbreak,
  title={Prisonbreak: Jailbreaking large language models with fewer than twenty-five targeted bit-flips},
  author={Coalson, Zachary and Woo, Jeonghyun and Chen, Shiyang and Sun, Yu and Yang, Lishan and Nair, Prashant and Fang, Bo and Hong, Sanghyun},
  journal={arXiv preprint arXiv:2412.07192},
  year={2024}
}

@article{xu2025silentstriker,
  title={Silentstriker: Toward stealthy bit-flip attacks on large language models},
  author={Xu, Haotian and Peng, Qingsong and Shi, Jie and Zheng, Huadi and Li, Yu and Zhuo, Cheng},
  journal={arXiv preprint arXiv:2509.17371},
  year={2025}
}

@misc{merity2016pointer,
      title={Pointer Sentinel Mixture Models},
      author={Stephen Merity and Caiming Xiong and James Bradbury and Richard Socher},
      year={2016},
      eprint={1609.07843},
      archivePrefix={arXiv},
      primaryClass={cs.CL}
}

@inproceedings{Dua2019DROP,
  author={Dheeru Dua and Yizhong Wang and Pradeep Dasigi and Gabriel Stanovsky and Sameer Singh and Matt Gardner},
  title={  {DROP}: A Reading Comprehension Benchmark Requiring Discrete Reasoning Over Paragraphs},
  booktitle={Proc. of NAACL},
  year={2019}
}

@article{cobbe2021gsm8k,
  title={Training Verifiers to Solve Math Word Problems},
  author={Cobbe, Karl and Kosaraju, Vineet and Bavarian, Mohammad and Chen, Mark and Jun, Heewoo and Kaiser, Lukasz and Plappert, Matthias and Tworek, Jerry and Hilton, Jacob and Nakano, Reiichiro and Hesse, Christopher and Schulman, John},
  journal={arXiv preprint arXiv:2110.14168},
  year={2021}
}

@article{2017arXivtriviaqa,
       author = {{Joshi}, Mandar and {Choi}, Eunsol and {Weld},
                 Daniel and {Zettlemoyer}, Luke},
        title = "{triviaqa: A Large Scale Distantly Supervised Challenge Dataset for Reading Comprehension}",
      journal = {arXiv e-prints},
         year = 2017,
          eid = {arXiv:1705.03551},
        pages = {arXiv:1705.03551},
archivePrefix = {arXiv},
       eprint = {1705.03551},
}

@misc{deepseekai2025deepseekr1incentivizingreasoningcapability,
      title={DeepSeek-R1: Incentivizing Reasoning Capability in LLMs via Reinforcement Learning}, 
      author={DeepSeek-AI},
      year={2025},
      eprint={2501.12948},
      archivePrefix={arXiv},
      primaryClass={cs.CL},
      url={https://arxiv.org/abs/2501.12948}, 
}

@misc{openai_gpt5_2026,
  author       = {{OpenAI}},
  title        = {{GPT-5}},
  howpublished = {\url{https://www.openai.com/}},
  year         = {2026},
  note         = {Large language model, accessed January 28, 2026}
}

@inproceedings{razavi2016flip,
  title={Flip feng shui: Hammering a needle in the software stack},
  author={Razavi, Kaveh and Gras, Ben and Bosman, Erik and Preneel, Bart and Giuffrida, Cristiano and Bos, Herbert},
  booktitle={25th USENIX Security Symposium (USENIX Security 16)},
  pages={1--18},
  year={2016}
}

@inproceedings{qiao2016new,
  title={A new approach for rowhammer attacks},
  author={Qiao, Rui and Seaborn, Mark},
  booktitle={2016 IEEE international symposium on hardware oriented security and trust (HOST)},
  pages={161--166},
  year={2016},
  organization={IEEE}
}

@inproceedings{gruss2016rowhammer,
  title={Rowhammer. js: A remote software-induced fault attack in javascript},
  author={Gruss, Daniel and Maurice, Cl{\'e}mentine and Mangard, Stefan},
  booktitle={International conference on detection of intrusions and malware, and vulnerability assessment},
  pages={300--321},
  year={2016},
  organization={Springer}
}

@article{seaborn2015exploiting,
  title={Exploiting the DRAM rowhammer bug to gain kernel privileges},
  author={Seaborn, Mark and Dullien, Thomas},
  journal={Black Hat},
  volume={15},
  number={71},
  pages={2},
  year={2015},
  publisher={USA}
}

@inproceedings{cojocar2019exploiting,
  title={Exploiting correcting codes: On the effectiveness of ecc memory against rowhammer attacks},
  author={Cojocar, Lucian and Razavi, Kaveh and Giuffrida, Cristiano and Bos, Herbert},
  booktitle={2019 IEEE Symposium on Security and Privacy (SP)},
  pages={55--71},
  year={2019},
  organization={IEEE}
}

@inproceedings{frigo2020trrespass,
  title={TRRespass: Exploiting the many sides of target row refresh},
  author={Frigo, Pietro and Vannacc, Emanuele and Hassan, Hasan and Van Der Veen, Victor and Mutlu, Onur and Giuffrida, Cristiano and Bos, Herbert and Razavi, Kaveh},
  booktitle={2020 IEEE Symposium on Security and Privacy (SP)},
  pages={747--762},
  year={2020},
  organization={IEEE}
}

@article{mukundan2013understanding,
  title={Understanding and mitigating refresh overheads in high-density DDR4 DRAM systems},
  author={Mukundan, Janani and Hunter, Hillery and Kim, Kyu-hyoun and Stuecheli, Jeffrey and Mart{\'\i}nez, Jos{\'e} F},
  journal={ACM SIGARCH Computer Architecture News},
  volume={41},
  number={3},
  pages={48--59},
  year={2013},
  publisher={ACM New York, NY, USA}
}

@inproceedings{jattke2024zenhammer,
  title={$\{$ZenHammer$\}$: Rowhammer Attacks on $\{$AMD$\}$ Zen-based Platforms},
  author={Jattke, Patrick and Wipfli, Max and Solt, Flavien and Marazzi, Michele and B{\"o}lcskei, Matej and Razavi, Kaveh},
  booktitle={33rd USENIX Security Symposium (USENIX Security 24)},
  pages={1615--1633},
  year={2024}
}

@inproceedings{jattke2025mcsee,
  title={$\{$McSee$\}$: Evaluating Advanced Rowhammer Attacks and Defenses via Automated $\{$DRAM$\}$ Traffic Analysis},
  author={Jattke, Patrick and Marazzi, Michele and Solt, Flavien and Wipfli, Max and Gloor, Stefan and Razavi, Kaveh},
  booktitle={34th USENIX Security Symposium (USENIX Security 25)},
  pages={5621--5640},
  year={2025}
}

@inproceedings{gloor2025refault,
  title={REFault: A Fault Injection Platform for Rowhammer Research on DDR5 Memory},
  author={Gloor, Stefan and Jattke, Patrick and Razavi, Kaveh},
  booktitle={Proceedings of the Microarchitecture Security Conference},
  year={2025}
}

@inproceedings{meyer2026phoenix,
  title={Phoenix: Rowhammer Attacks on DDR5 with Self-Correcting Synchronization},
  author={Meyer, Diego and Jattke, Patrick and Marazzi, Michele and Qazi, Salman and Moghimi, Daniel and Razavi, Kaveh},
  booktitle={Proceedings of the 2026 IEEE Symposium on Security Privacy (S\&P), San Francisco, CA, USA},
  year={2026}
}

@article{lin2024awq,
  title={Awq: Activation-aware weight quantization for on-device llm compression and acceleration},
  author={Lin, Ji and Tang, Jiaming and Tang, Haotian and Yang, Shang and Chen, Wei-Ming and Wang, Wei-Chen and Xiao, Guangxuan and Dang, Xingyu and Gan, Chuang and Han, Song},
  journal={Proceedings of machine learning and systems},
  volume={6},
  pages={87--100},
  year={2024}
}

@incollection{gholami2022survey,
  title={A survey of quantization methods for efficient neural network inference},
  author={Gholami, Amir and Kim, Sehoon and Dong, Zhen and Yao, Zhewei and Mahoney, Michael W and Keutzer, Kurt},
  booktitle={Low-power computer vision},
  pages={291--326},
  year={2022},
  publisher={Chapman and Hall/CRC}
}

@article{zhu2024survey,
  title={A survey on model compression for large language models},
  author={Zhu, Xunyu and Li, Jian and Liu, Yong and Ma, Can and Wang, Weiping},
  journal={Transactions of the Association for Computational Linguistics},
  volume={12},
  pages={1556--1577},
  year={2024},
  publisher={MIT Press 255 Main Street, 9th Floor, Cambridge, Massachusetts 02142, USA~…}
}

@article{dettmers2022gpt3,
  title={Gpt3. int8 (): 8-bit matrix multiplication for transformers at scale},
  author={Dettmers, Tim and Lewis, Mike and Belkada, Younes and Zettlemoyer, Luke},
  journal={Advances in neural information processing systems},
  volume={35},
  pages={30318--30332},
  year={2022}
}

@article{frantar2022gptq,
  title={Gptq: Accurate post-training quantization for generative pre-trained transformers},
  author={Frantar, Elias and Ashkboos, Saleh and Hoefler, Torsten and Alistarh, Dan},
  journal={arXiv preprint arXiv:2210.17323},
  year={2022}
}

@article{kalamkar2019study,
  title={A study of BFLOAT16 for deep learning training},
  author={Kalamkar, Dhiraj and Mudigere, Dheevatsa and Mellempudi, Naveen and Das, Dipankar and Banerjee, Kunal and Avancha, Sasikanth and Vooturi, Dharma Teja and Jammalamadaka, Nataraj and Huang, Jianyu and Yuen, Hector and others},
  journal={arXiv preprint arXiv:1905.12322},
  year={2019}
}

@inproceedings{wolf2020transformers,
  title={Transformers: State-of-the-art natural language processing},
  author={Wolf, Thomas and Debut, Lysandre and Sanh, Victor and Chaumond, Julien and Delangue, Clement and Moi, Anthony and Cistac, Pierric and Rault, Tim and Louf, Remi and Funtowicz, Morgan and others},
  booktitle={Proceedings of the 2020 conference on empirical methods in natural language processing: system demonstrations},
  pages={38--45},
  year={2020}
}

@inproceedings{chen2017zoo,
  title={Zoo: Zeroth order optimization based black-box attacks to deep neural networks without training substitute models},
  author={Chen, Pin-Yu and Zhang, Huan and Sharma, Yash and Yi, Jinfeng and Hsieh, Cho-Jui},
  booktitle={Proceedings of the 10th ACM workshop on artificial intelligence and security},
  pages={15--26},
  year={2017}
}

@article{gu2019badnets,
  title={Badnets: Evaluating backdooring attacks on deep neural networks},
  author={Gu, Tianyu and Liu, Kang and Dolan-Gavitt, Brendan and Garg, Siddharth},
  journal={Ieee Access},
  volume={7},
  pages={47230--47244},
  year={2019},
  publisher={IEEE}
}

@article{szegedy2013intriguing,
  title={Intriguing properties of neural networks},
  author={Szegedy, Christian and Zaremba, Wojciech and Sutskever, Ilya and Bruna, Joan and Erhan, Dumitru and Goodfellow, Ian and Fergus, Rob},
  journal={arXiv preprint arXiv:1312.6199},
  year={2013}
}

@inproceedings{liu2018trojaning,
  title={Trojaning attack on neural networks},
  author={Liu, Yingqi and Ma, Shiqing and Aafer, Yousra and Lee, Wen-Chuan and Zhai, Juan and Wang, Weihang and Zhang, Xiangyu},
  booktitle={25th Annual Network And Distributed System Security Symposium (NDSS 2018)},
  year={2018},
  organization={Internet Soc}
}

@article{li2024backdoorllm,
  title={Backdoorllm: A comprehensive benchmark for backdoor attacks on large language models},
  author={Li, Yige and Huang, Hanxun and Zhao, Yunhan and Ma, Xingjun and Sun, Jun},
  journal={arXiv e-prints},
  pages={arXiv--2408},
  year={2024}
}

@inproceedings{yao2020deephammer,
  title={$\{$DeepHammer$\}$: Depleting the intelligence of deep neural networks through targeted chain of bit flips},
  author={Yao, Fan and Rakin, Adnan Siraj and Fan, Deliang},
  booktitle={29th USENIX Security Symposium (USENIX Security 20)},
  pages={1463--1480},
  year={2020}
}
\end{document}